# Towards Zero-Overhead Disambiguation of Deep Priority Conflicts

Luis Eduardo de Souza Amorim[a], Michael J. Steindorfer[a], and Eelco Visser[a]

a   Delft University of Technology, Delft, The Netherlands

**Abstract**

**Context** Context-free grammars are widely used for language prototyping and implementation. They allow formalizing the syntax of domain-specific or general-purpose programming languages concisely and declaratively. However, the natural and concise way of writing a context-free grammar is often ambiguous. Therefore, grammar formalisms support extensions in the form of *declarative disambiguation rules* to specify operator precedence and associativity, solving ambiguities that are caused by the subset of the grammar that corresponds to expressions.

   **Inquiry** Implementing support for declarative disambiguation within a parser typically comes with one or more of the following limitations in practice: a lack of parsing performance, or a lack of modularity (i.e., disallowing the composition of grammar fragments of potentially different languages). The latter subject is generally addressed by scannerless generalized parsers. We aim to equip scannerless generalized parsers with novel disambiguation methods that are inherently performant, without compromising the concerns of modularity and language composition.

   **Approach** In this paper, we present a novel low-overhead implementation technique for disambiguating deep associativity and priority conflicts in scannerless generalized parsers with lightweight data-dependency.

   **Knowledge** Ambiguities with respect to operator precedence and associativity arise from combining the various operators of a language. While *shallow conflicts* can be resolved efficiently by one-level tree patterns, *deep conflicts* require more elaborate techniques, because they can occur arbitrarily nested in a tree. Current state-of-the-art approaches to solving deep priority conflicts come with a severe performance overhead.

   **Grounding** We evaluated our new approach against state-of-the-art declarative disambiguation mechanisms. By parsing a corpus of popular open-source repositories written in Java and OCaml, we found that our approach yields speedups of up to 1.73 x over a grammar rewriting technique when parsing programs with deep priority conflicts—with a modest overhead of 1 % to 2 % when parsing programs without deep conflicts.

   **Importance** A recent empirical study shows that deep priority conflicts are indeed wide-spread in real-world programs. The study shows that in a corpus of popular OCaml projects on Github, up to 17 % of the source files contain deep priority conflicts. However, there is no solution in the literature that addresses efficient disambiguation of deep priority conflicts, with support for modular and composable syntax definitions.

## The Art, Science, and Engineering of Programming

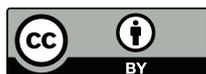

**Towards Zero-Overhead Disambiguation of Deep Priority Conflicts**

# 1 Introduction

Context-free grammars have been established as the main formalism for concisely describing the syntax of programming languages (e.g., in reference manuals). Yet, context-free grammar definitions still cause problems when used to generate parsers in practice. On the one hand, a parser generator may expect a deterministic grammar that fits a certain grammar class, such as LL or LR. On the other hand, natural and concise context-free grammars may be inherently ambiguous, more specifically when considering the subset of the grammar that defines expressions and operators.

Mainstream parser generators, such as YACC [13], extend their grammar formalism with declarative constructs for disambiguation, allowing users to specify the precedence and associativity of operators. In order to not compromise performance, under the hood, YACC's translation of disambiguation rules highly depends on specific characteristics of LR parsing technology—exploiting LR shift/reduce conflicts—rather than building upon a generalized solution. Furthermore, YACC does not support the composition of modular grammar fragments (of potentially different languages).

In contrast, the SDF2 syntax definition formalism [25] allows modular and composable language specifications, providing mechanisms for declaratively specifying operator precedence and associativity [15, 22]. The semantics for SDF2 disambiguation is parsing independent, but it only addresses ambiguities that are caused by so called *shallow conflicts*, i.e., conflicts that can be solved by checking whether a certain parse node is a direct descendant of another node in the parse tree. However, some ambiguities that occur in expressions can only be solved by checks in the final tree of unbounded depth [3, 19]. Such ambiguities are caused by *deep priority conflicts*.

Several approaches have been proposed to solve deep priority conflicts. Many of these approaches are based on grammar transformations and thus are parser independent [2, 3, 19]. However, those techniques typically result in large unambiguous grammars, which may impact on the performance of the parser and be somewhat inefficient, as considerable parts of the grammar are not exercised at runtime, even after parsing many programs [20].

Alternative solutions of so called data-dependent grammars [5] postpone solving of priority conflicts to parse time. Data-dependent grammars are context-free grammars, extended with arbitrary computations, parameters, variable binding, and constraints that can be evaluated at parse time [12]. Data dependent grammars that address disambiguation of priority conflicts can be generated from a context-free grammar with disambiguation constructs, but they are fairly complex and arguably hard to read and understand. The additional complexity comes from the bindings and constraints, and by the fact that data can be propagated "downwards" and "upwards" at parse time, when building the parse tree. Finally, data-dependent grammars have not yet been generalized to solve frequent types of deep priority conflicts such as longest match or the well known dangling else problem.

This paper proposes a different solution to disambiguate deep priority conflicts at parse time based on *data-dependent contextual grammars*. Our approach relies on lightweight data dependency that does not require arbitrary computation nor variable bindings at parse time, and instead expresses disambiguation in terms of set-algebraic

**Luis Eduardo de Souza Amorim, Michael J. Steindorfer, and Eelco Visser**

operations that can be implemented scoped and efficiently. The contributions of the paper are:

- We define a lightweight data-dependent extension of the scannerless generalized LR parsing algorithm for disambiguating deep priority conflicts.
- We show that this data-dependent extension yields the same disambiguation as contextual grammars, an approach that solves deep priority conflicts through grammar rewriting.
- We compare the performance of disambiguation strategies and show that our lightweight data-dependent disambiguation is up to 1.73 x faster when parsing programs with deep priority conflicts and has very low overhead for programs without deep priority conflicts.

We implemented our solution as a modified parser generator for SDF3 [26], which supports modular and composable syntax definitions. Furthermore, we evaluated our approach using OCaml, an expression-based language that contains many deep priority conflicts; and Java, a statement-based language that contains a small number of deep priority conflicts. Given that deep priority conflicts may occur in about one in five real-world programs for OCaml [20], we provide a technique that supports efficient disambiguation of such conflicts, showing that for programs without conflicts, our approach has a modest overhead of 1 % to 2 % on parsing time.

The paper is organized as follows: Section 2 details background information on declarative disambiguation and deep priority conflicts. Section 3 describes data-dependent contextual grammars. Next, in Section 4 we evaluate our approach by comparing to disambiguation techniques that rely on grammar transformations. Finally, we discuss related work in Section 5, before concluding.

## 2 Disambiguating Priority Conflicts

We start by presenting the notation for grammars, grammar productions and parse trees that will be used throughout the paper. The remainder of this section then discusses background on the issue of deep priority conflicts and a grammar rewriting technique, contextual grammars [19], that addresses the resolution of such conflicts.

### 2.1 Notation

**Grammars**  A context-free grammar G can be formally defined as a tuple $(\Sigma, N, P)$ with the set $\Sigma$ representing the terminal symbols; the set $N$ consisting of the non-terminal symbols defined in the grammar; and the set $P$ representing the productions. When not mentioned, we adopted the letter $A$ to represent arbitrary non-terminals; the letter X to represent a symbol from $\Sigma \cup N$; and Greek letters $\alpha$, $\beta$ or $\gamma$ to represent a symbol in $(\Sigma \cup N)^*$, also known as sentential forms.

**Productions**  We use the same notation for productions as in the syntax definition formalism SDF3 [26]. A production in a grammar G has the form `A = `$\alpha$ or `A.C = `$\alpha$, where `C` represents a constructor. SDF3 productions may have constructors to specify the name of the abstract syntax tree node constructed when imploding the parse tree. A non-terminal and a constructor uniquely identify a production, i.e., a production `A.C = α` may also be referred as `A.C`. Note that SDF3 constructors are orthogonal to our approach.

**Parse Trees** A production `A.C = X`$_1$` ... X`$_n$ may be used to construct a tree of form `[A.C = t`$_1$` ... t`$_n$`]`, with the subtree `t`$_i$ being defined by the symbol `X`$_i$. Explicit subtrees are indicated by their productions using nested square brackets, whereas arbitrary subtrees and terminal elements that occur as leaves do not require brackets. For example, the tree `[Exp.Add = e`$_1$` + [Exp.Mul = e`$_2$` * e`$_3$`]]` constructed with a production `Exp.Add = Exp "+" Exp` has an arbitrary subtree `e`$_1$ as its leftmost subtree, and a rightmost explicit subtree defined by the production `Exp.Mul = Exp "*" Exp`, with arbitrary subtrees `e`$_2$ and `e`$_3$.

## 2.2 Background on Deep Priority Conflicts

In context-free grammars of programming languages, ambiguities are often caused by the subset of the language that contains expressions and operators. To address this issue, grammar formalisms used in practice support the specification of declarative disambiguation rules to define operator precedence and associativity among the grammar productions. E.g., in SDF3, a grammar production can have an annotation—either `left`, `right`, or `non-assoc`—to specify its associativity. SDF3 also supports context-free priorities, which form a partial order, defining a priority relation between productions. E.g., the disambiguation rule `Exp.Add > Exp.If` defines that addition has a higher priority than conditional expressions.

The most common ambiguities from expression grammars involve the direct combination of operators with different priorities. These ambiguities are caused by so-called *shallow priority conflicts* and can be efficiently solved by subtree filtering [15], i.e., disallowing certain kinds of trees to occur as a direct child of others.

A small but complicated-to-solve subset of ambiguities is caused by *deep priority conflicts*. Unlike shallow conflicts, deep priority conflicts cannot be filtered by observing the direct parent-child relationship of nodes within a parse tree. Deep priority conflicts can occur arbitrarily nested (i.e., in unbounded depth) in a parse tree. In general, deep conflicts can occur when a low-priority operator shadows a nested higher priority operator on the left- or rightmost edges along a sub-tree [3, 20]. Deep priority conflicts are commonly found in the expression parts of grammars of ML-like languages. A recent empirical pilot study suggests that up to 17 % of OCaml source files originating from popular projects on Github do contain deep priority conflicts [20], raising the question how such conflicts can be disambiguated efficiently.

Deep priority conflicts are categorized in three classes, according to their nature [19]:

**Listing 1** Expression grammars with various archetypes of deep priority conflicts.

**(a)** Operator-Style Conflict

```
context-free syntax

  Exp.If  = "if" "(" Exp ")" Exp
  Exp.Add = Exp "+" Exp {left}
  Exp.Int = INT

context-free priorities

  Exp.Add > Exp.If

causes conflict in sentence

  e₁ + if(e₂) e₃ + e₄

with interpretations

  e₁ + if(e₂) (e₃ + e₄)

  (e₁ + if(e₂) e₃) + e₄
```

**(b)** Dangling-Else Conflict

```
context-free syntax

  Exp.If     = "if" "(" Exp ")" Exp
  Exp.IfElse = "if" "(" Exp ")" Exp "else" Exp
  Exp.Int    = INT

causes conflict in sentence

  if(e₁) if(e₂) e₃ else if(e₄) e₅ else e₆

with interpretations

  if(e₁) (if(e₂) e₃ else (if(e₄) e₅ else e₆))

  if(e₁) (if(e₂) e₃ else (if(e₄) e₅)) else e₆

  if(e₁) (if(e₂) e₃) else (if(e₄) e₅ else e₆)
```

**(c)** Longest-Match Conflict

```
context-free syntax

  Exp.Match   = "match" Exp "with" Pat+
  Pat.Pattern = ID "->" Exp
  Exp.Int     = INT

causes conflict in sentence

  match e₁ with id -> match e₂ with p₁ p₂

with interpretations

  match e₁ with id -> (match e₂ with p₁ p₂)

  match e₁ with id -> (match e₂ with p₁) p₂
```

**Towards Zero-Overhead Disambiguation of Deep Priority Conflicts**

**Operator-Style Conflicts** Operator-style conflicts involve two operators: 1) a prefix operator[1] with lower priority, and 2) a postfix or infix operator with higher priority.[2] Listing 1a contains a minimal grammar example, illustrating an operator-style conflict. In our case, it involves an addition expression that has higher priority than the conditional expression. Parsing the example sentence on line 13 causes an ambiguity due to a deep priority conflict and yields two possible interpretations (lines 17 and 19). In the example, the first instance represents the supposedly correct interpretation, since the addition to the left of the conditional expression extends as far as possible. The second and incorrect interpretation cannot be filtered by checking the direct parent-child relation of parse nodes, since the first addition expression shadows that the conditional (prefix operator with lower-priority) occurs indirectly nested at the rightmost position of the second addition (infix operator with higher-priority).

**Dangling-Else Conflicts** Dangling-else describes a pattern for a deep priority conflict involving two productions that share the same prefix or suffix, where the shorter production is (left or right) recursive. Listing 1b illustrates a conflict involving the `Exp.If` and `Exp.IfElse` productions. For the sentence on line 9, a parser cannot decide where the else branches should be connected. Note that the first interpretation (line 13) is supposedly the correct one, where the else branches are connected to the closest if-expressions.

**Longest-Match Conflicts** Another type of deep priority conflict involves indirectly nested lists [19]. The example grammar in Listing 1c defines `Exp.Match` expressions ending with a list of patterns. However, match expressions can themselves occur at the end of a pattern. E.g., for the sentence in line 9, the parser cannot decide whether the pattern $p_2$ belongs to the list of the `match` $e_1$ (cf. line 15) or the list of the `match` $e_2$ expression (cf. line 13). The first interpretation should be preferred if the list construct (`Pat+`) itself follows longest match.

**2.3 Disambiguating Deep Priority Conflicts with Contextual Grammars**

Many disambiguation approaches achieve independence from a particular parsing technology by relying on grammar rewriting (i.e., transforming an ambiguous context-free grammar into a context-free grammar that does not contain any priority conflicts). In the following, we discuss the disambiguation approach of contextual grammars [19] as a representative example for rewriting-based disambiguation strategies, since our contribution builds upon it. (An extensive discussion and comparison of related work on the subject of disambiguation can be found in Section 5).

---

[1] We consider the definition of operators used in [19]: prefix operators are defined by *right* recursive productions, postfix operators by *left* recursive productions and infix operators by productions that are both *left* and *right* recursive.
[2] Operator-style conflicts may also involve lower priority postfix operators, but these are uncommon, as postfix operators usually have higher priority in most programming languages.

**Luis Eduardo de Souza Amorim, Michael J. Steindorfer, and Eelco Visser**

■ **Listing 2** Contextual grammar that solves an operator-style deep priority conflict involving if and addition expressions in ML-like languages.

```
1  context-free syntax
2
3    Exp.If  = "if" "(" Exp ")" Exp
4    Exp.Add = Exp^{Exp.If} "+" Exp {left}
5    Exp.Int = INT
6
7    Exp^{Exp.If}.Add = Exp^{Exp.If} "+" Exp^{Exp.If} {left}
8    Exp^{Exp.If}.Int = INT
9
10 context-free priorities
11
12    Exp.Add > Exp.If
13
14 uniquely parses sentence
15
16    e_1 + if(e_2) e_3 + e_4
17
18 with interpretation
19
20    e_1 + if(e_2) (e_3 + e_4)
```

Contextual grammars [19] are context-free grammars that can be used to solve deep priority conflicts. Under the hood, contextual grammars express invalid parse-tree patterns with respect to the disambiguation rules defined in the grammar. These patterns can be deeply matched to filter trees that would cause an ambiguity. The deep pattern matches do not occur at parse-time, but rather are implemented as a grammar transformation. A "black-list" of forbidden patterns, represented by so-called *contextual tokens* drives the recursive rewriting algorithm, restricting which parse trees a production may produce along the (leftmost or rightmost) positions of sub-trees.

**Recursive Rewriting by Example**  Listing 2 illustrates an example for a contextual grammar that solves the operator-style conflicts of Listing 1a. A grammar transformation recursively rewrites the grammar and adds new productions for the symbol Exp$^{Exp.If}$ (lines 7– 8). The rewriting propagates the contextual tokens to all leftmost and rightmost non-terminals of the newly added production. By using the tokens to create new non-terminal symbols that implement filters, the rewriting avoids the construction of invalid trees. According to this grammar, the example sentence in line 16 now can be unambiguously parsed as shown in line 20. The invalid sentence ($e_1$ + if($e_2$) $e_3$) + $e_4$ cannot be parsed anymore, because the addition … + $e_4$ that is parsed with Exp.Add must not have an if-expression on the rightmost position of the first addition $e_1$ + if($e_2$) $e_3$.

One issue with the rewriting is that the propagation of constraints might result in many additional productions in the final contextual grammar, as productions of the

**Towards Zero-Overhead Disambiguation of Deep Priority Conflicts**

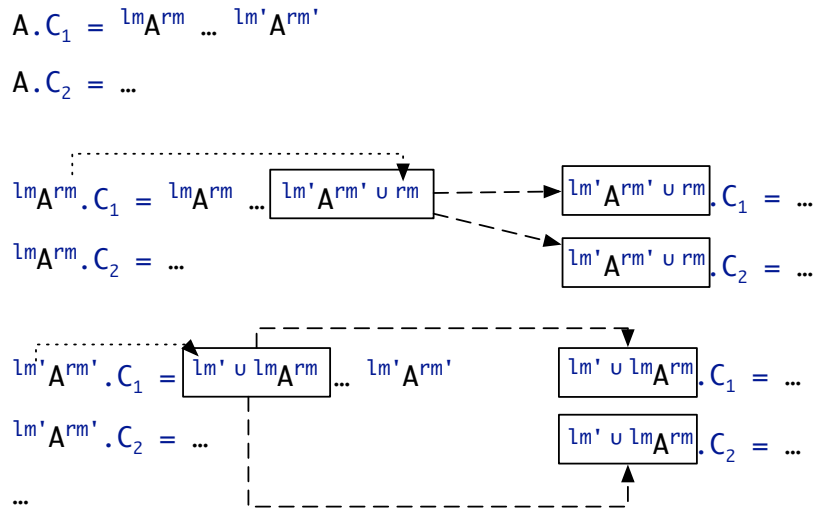

**Figure 1** Generalized recursive propagation of contextual tokens in contextual grammars.

original grammar need to be duplicated (recursively) for each new contextual symbol. While the previous example is relatively concise, the general case is not.

**Recursive Rewriting in General** Formally, a contextual symbol $^{lm}A^{rm}$ is a regular nonterminal $A$ that is uniquely identified by the tuple $(lm, A, rm)$, where $lm$ and $rm$ are sets containing contextual tokens. For brevity we omit $lm$ or $rm$ respectively when the set is empty. The set $lm$ stores unique references to productions that are not allowed to occur in any, possibly deeply nested, leftmost node of the tree defined by the contextual symbol $^{lm}A^{rm}$. Similarly, the productions referenced in the set $rm$ cannot be used to construct any, possibly deeply nested, rightmost node of the tree defined by $^{lm}A^{rm}$. Finally, the tree for the symbol $A$ itself cannot be constructed using any of the productions referenced in $lm$ and $rm$.

Figure 1 highlight the general case of how productions are recursively rewritten. The starting point are the first two productions $A.C_1$ and $A.C_2$, where the production $A.C_1$ contains deep priority conflicts, as indicated by the contextual symbols on the right-hand side of the production. For each unique contextual symbol, the productions for the symbol $A$ need to be duplicated excluding the productions in the sets $lm$ and $rm$, while propagating the contextual tokens accordingly. In the second pair of rules in the grammar, new productions are created for the symbol $^{lm}A^{rm}$ assuming that $C_1$ and $C_2$ are not in the sets $lm$ and $rm$, and the set $rm$ is propagated to the rightmost symbol of that rule (cf. dotted arrow). When propagating the set containing the rightmost contextual tokens $rm$ to the rightmost symbol of this rule, a new unique contextual symbol $^{lm'}A^{rm' \cup rm}$ is generated, causing a ripple effect: new productions need to be recursively generated for the new symbol as well (cf. the dashed arrows). The same ripple effect might occur for the symbol $^{lm' \cup lm}A^{rm'}$ and for any other unique contextual symbol resulting from the propagation of contexts.

Contextual grammars can correctly disambiguate all previously discussed conflicts, however at a high cost. For expression-based languages with many deep priority conflicts—such as OCaml—the grammar can get about three times bigger [19]. In the case of contextual grammars, the duplication is necessary to solve deep conflicts, however, many productions are not exercised in practice, even after parsing a large set of programs [20]. The duplication directly introduces a performance penalty, causing larger parse tables, and longer parse times in practice.

Our aim is to avoid the blow-up in productions caused by grammar transformations, without giving up the correctness properties guaranteed by contextual grammars. In the next section we illustrate how the underlying concepts of contextual grammars can be repurposed to disambiguate deep priority conflicts efficiently at parse time.

## 3 Data-dependent Contextual Grammars

In this section, we focus on declarative disambiguation techniques that are more general than, for example, YACC's approach, in order to support modular and composable syntax definitions. In particular, we illustrate how low-overhead disambiguation can be implemented in SDF3 [26] with a scannerless generalized LR parser (SGLR) [24].

Figure 2 highlights the different stages in the context of parser generation using SDF3 and parsing in SGLR. First, a *normalized*[3] SDF3 grammar is first transformed by recursive rewriting into a contextual grammar, which contains additional productions to remove deep priority conflicts (cf. Section 2.3). Second, the parse table generator produces a parse table given the contextual grammar, solving shallow priority conflicts directly when constructing the table, by filtering *goto*-transitions according to the priorities specified in the grammar [22]. Afterwards, the SGLR parser uses the generated parse table for processing arbitrary input programs. The parser may use other disambiguation mechanisms at parse time to address, for example, ambiguities in the lexical syntax using reject productions [24]. The SGLR parser returns a compact representation of a parse forest that contains all trees that were derived when parsing an input program. As a final step, a (post-parse) disambiguator may still remove invalid trees from the parse forest according to given constraints.

According to Figure 2, we can identify four different stages when disambiguation of priority conflicts can occur: (1) before parse table generation, (2) at parser generation, (3) at parse time, and (4) after parsing. Post-parse disambiguation is conceptually the most expensive approach, since ambiguities can grow exponentially with the size of the input [10]. Disambiguation should preferably occur in the first three identified stages of disambiguation (i.e., avoiding the construction of invalid trees beforehand). Nevertheless, disambiguating early in the pipeline does not necessarily guarantee the best performance either, as we will discuss next.

---

[3] SDF3 grammars are normalized to handle lexical and context free syntax declarations, derive additional productions for symbols that represent lists or optionals, insert optional layout in between context-free symbols, and expand priority groups and chains.

**Towards Zero-Overhead Disambiguation of Deep Priority Conflicts**

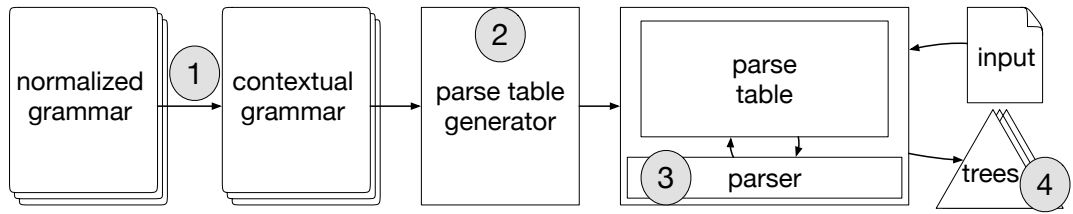

**Figure 2** Parsing a program with a scannerless generalized parser, and the times when disambiguation might occur.

**When to Disambiguate (Deep) Priority Conflicts?** Disambiguating priority conflicts by grammar rewriting occurs before parse table generation. Rewriting techniques have the advantage that the remainder of the parser generation and parsing pipeline can operate oblivious of priority conflicts. Especially for resolving deep priority conflicts, grammar rewriting unfortunately adds many productions for forbidding conflicting patterns and may result in large grammars that negatively impact performance.

Disambiguating conflicts during parse table generation does not require any grammar rewriting and can be achieved by modifying the LR parse table generator. In a scannerless parser, disambiguation at parse table generation can only resolve shallow priority conflicts, requiring that deep priority conflicts are addressed earlier or later [19].

As noted previously, using post-parse disambiguation filters to solve priority conflicts can be inefficient in practice, because the number of ambiguities in expressions can grow exponentially with the size of an expression. Hence, a post-parse filter would have to traverse a large number of trees in the parse forest to filter invalid trees.

According to the reasoning above, in the next sections we will explore a solution to disambiguate deep priority conflicts efficiently at parse time, while keeping the disambiguation of shallow conflicts when generating the parse table.

### 3.1 Disambiguation of Deep Conflicts with Lightweight Data Dependency

Data-dependent grammars [12] extend context-free grammars allowing parameterized non-terminals, variable binding, evaluation of constraints, and arbitrary computation at parse time. Data-dependent grammars can be translated into stack-based automata, i.e., push-down automata with environments to track data-dependent parsing states. For example, consider the productions:

```
Iter(n).Conc  = [n >= 1] Iter(n - 1) A
Iter(n).Empty = [n == 0] ε
```

In the example above, the non-terminal `Iter` is parameterized by an integer `n`, which indicates the length of the iteration over the non-terminal `A`. The constraint `[n >= 1]` is checked before trying to parse `Iter(n - 1) A`, i.e., if $n \geq 1$ the first production is used, otherwise, the second.

Purely data dependent grammars are powerful enough to disambiguate priority conflicts of grammars for programming languages at parse time [5]. They allow resolution of possible priority conflicts in the grammar by means of constraints that forbid the creation of invalid trees. Nevertheless, just relying on data-dependency might negatively impact the performance of the parser, especially when parsing files that are free of priority conflicts. For that reason, we selectively use a lightweight form of data-dependency to solely solve deep priority conflicts, without requiring variable bindings or arbitrary computations at parse time.

**Leveraging Data-Dependency to Avoid Duplicating Productions**  Contextual grammars (cf. Section 2.3) can be treated as pure context-free grammars, if we consider that every unique contextual symbol specifies a new non-terminal. Transforming a contextual grammar into a context-free one, occurs by duplicating productions (recursively) for each unique contextual symbol.

Without duplicating the productions, a contextual grammar would have exactly the same shape and number of productions as the original grammar, since contextual symbols consist of essentially annotated non-terminals that originate from an analysis phase. Without rewriting, the grammar itself is still ambiguous, but the inferred contextual tokens that occur in the grammar can be reused to solve deep priority conflicts at parse time.

**Bottom-up Constraint Aggregation instead of Top-Down Rewriting**  Instead of propagating the constraints in the form of contextual tokens in the grammar productions, which may result in new contextual symbols and consequently new productions, we propagate the data to which the constraints are applied at parse time. Since the SGLR parser constructs trees bottom-up, we propagate the information about the productions used to construct the possibly nested leftmost and rightmost nodes of a tree bottom-up as contextual tokens during tree construction. Each node of the parse tree of the adapted data-dependent SGLR parser contains two additional sets that indicate the productions used to construct its leftmost and rightmost (nested) subtrees, respectively. For every node, the set representing the leftmost contextual tokens is the union of the the production used to construct the current node with the leftmost set of the leftmost direct child. Similarly, the set representing the rightmost contextual tokens is the union of the production used to construct the node itself with the rightmost set of contextual tokens of the rightmost direct child. Note that only productions that can cause deep priority conflicts are added to the sets of contextual tokens; the number of tokens propagated is significantly lower than the total number of productions, even for highly ambiguous grammars. (The largest number of contextual tokens—33—was required for OCaml, which contains a highly ambiguous expression grammar.)

**Data-Dependent Contextual Token Propagation by Example**  The tree in Figure 3a for the sentence `INT` `+` `if` `INT` was parsed using the data-dependent contextual grammar of Listing 3. Since `Exp.If` is the only production that appears in the contextual tokens in the grammar, it is the only token that needs to be propagated upwards. Because the if-expression occurs as a direct right subtree of the addition, only its rightmost set of contextual tokens is propagated upwards, discarding the leftmost set of tokens.

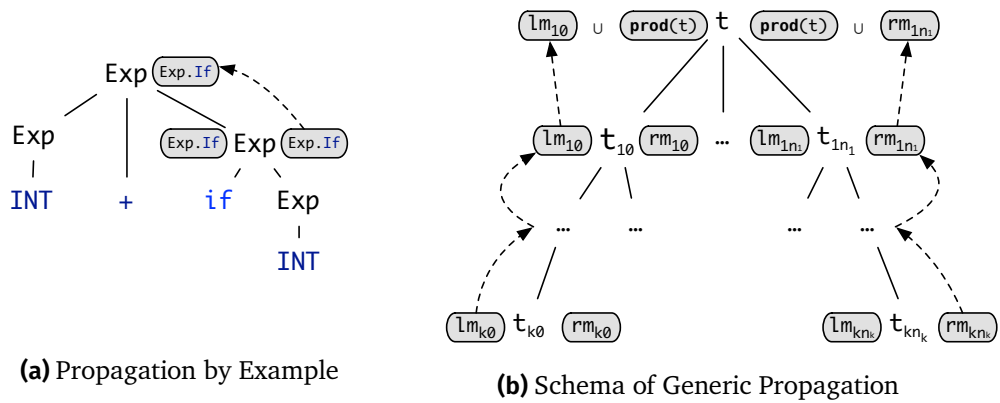

**(a)** Propagation by Example      **(b)** Schema of Generic Propagation

**Figure 3** Propagation of contextual tokens in contextual grammars with data-dependency.

**Listing 3** A (truncated) succinct syntax for a data-dependent contextual grammar with if-expressions, to be used for illustrating concise parse tree examples.

```
1 context-free syntax
2
3   Exp.If  = "if" Exp
4   Exp.Add = Exp^{Exp.If} "+" Exp {left}
5   Exp.Int = INT
6
7 context-free priorities
8
9   Exp.Add > Exp.If
```

**Data-Dependent Contextual Token Propagation in General** Consider the tree schema indicating the propagation of possible contextual tokens of Figure 3b. Assuming that the tree has depth $k$, the tokens will be propagated bottom-up through the leaves until reaching the root $t$. However, for a leaf node $t_{k0}$, its set of contextual tokens consist only of **prod**($t_{k0}$) (the production used to construct $t_{k0}$). As we will discuss in Section 3.3, we only propagate contextual tokens that occur in the contextual grammar, i.e., if **prod**($t_{k0}$) cannot cause a deep priority conflict, the set is in fact, an empty set. Thus, besides limiting the propagation to the depth of the trees being constructed, for grammars with few conflicts, only a small amount of data is actually propagated when constructing the tree.

**Customization of Parsing Algorithm** The algorithm for the data-dependent scannerless generalized LR parser requires only a few changes in the original SGLR algorithm shown in [24]. More specifically, the algorithm needs to propagate contextual tokens corresponding to the productions used to construct the leftmost and rightmost (possibly nested) subtrees (t.LeftmostTokens and t.RighmostTokens),[4] and to check

---

[4] In the original SGLR algorithm, creating a parse tree node consisted simply of applying a production to the trees collected when calculating the path for the reduce action. In the

**Listing 4** Pseudocode for the modified *DO-REDUCTIONS* and *CREATE-TREE-NODE* methods from the original SGLR, in the implementation of the data-dependent SGLR.

```
function DO-REDUCTIONS(Stack st, Production A.C = X_1...X_n)
  for each path from stack st to stack st_0 of length n} do
    List<Tree> [t_1,...,t_n] = the trees from the links in the path from st to st_0
    for each X_i such that X_i is a contextual symbol ^{lm}X^{rm} do
      if t_i.LeftmostTokens ∩ lm ≠ ∅ or t_i.RightmostTokens ∩ rm ≠ ∅ then
        return
      end if
    end for
    REDUCER(st_0, goto(state(st_0), A.C = X_1...X_n), A.C = X_1...X_n, [t_1,...,t_n])
  end for
end function
```

```
function CREATE-TREE-NODE(Production A.C = X_1...X_n, List<Tree> [t_1,...,t_n])
  Tree t = [A.C = t_1,...,t_n]
  t.LeftmostTokens = t_1.LeftmostTokens ∪ A.C
  t.RightmostTokens = t_n.RightmostTokens ∪ A.C
  return t
end function
```

the constraints when performing reduce actions. We show the pseudocode for the modified methods of the original SGLR in Listing 4. Note that because we leverage the analysis done by contextual grammars, our data-dependent SGLR algorithm can solve the same types of deep priority conflicts that can be solved by regular contextual grammars, i.e., operator-style, dangling else and longest match. Furthermore, because we propagate the data representing possible conflicts at parse time, and enforce the constraints when performing a reduce operation, the grammar does not require modifications that increase its number of productions.

### 3.2 Scannerless Generalized LR Parsing with Data-Dependent Disambiguation

To illustrate how our implementation of a data-dependent SGLR performs disambiguation at parse time, consider the scenario when parsing the input `INT + if INT + INT`, which contains an operator-style deep priority conflict, using the data-dependent contextual grammar shown previously. After parsing `INT + if INT`, the parser reaches a state with a shift/reduce conflict in the parse table shown in configuration (I) from Figure 4. Before this point, SGLR performs actions according to the parse table to construct the single stack shown in this configuration, with the links between states (represented by the boxes) containing the trees that have been created so far, or the terminal symbols that have been shifted.

Note that in this first configuration, when reaching the conflict in the parse table, a parser that uses the disambiguation mechanism from YACC (see Section 5) can

---

data-dependent algorithm, the sets of leftmost and rightmost subtrees need to be updated by propagating the information from the rightmost and leftmost direct subtrees.

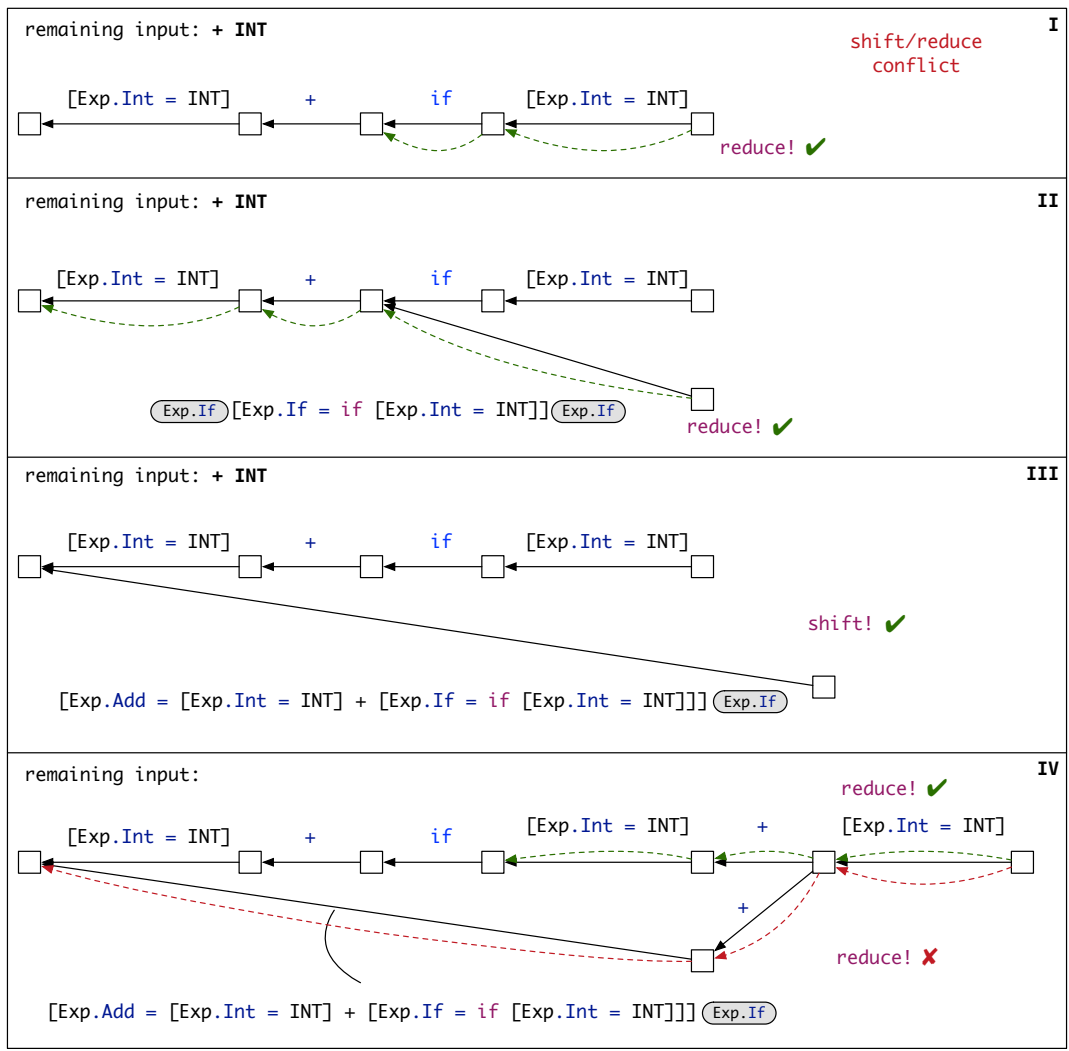

**Figure 4** The configurations of SGLR when solving a deep priority conflict when parsing program INT + if INT + INT.

make the decision of which action to take based on the next input token. In this case, the parser would choose shifting over reducing because the next token in the input is + and the addition has higher priority over the if expression. However, this approach of looking at the next input token does not extend to scannerless parsers or character-level grammars, since the parser operates on characters, and the character + might be preceded by layout, or could be the prefix of a different operator.

Thus, instead of making a decision at the configuration (I), a generalized parser such as SGLR performs both actions in pseudo-parallel, producing an ambiguity if both actions lead to a successful parse. First, SGLR performs all possible reduce actions, which may result in the creation of different stacks. That is, the parser continues by forking a new stack, adding a link to the original one, creating a graph structured stack. This occurs at the configuration (II), as a reduce action has been performed to construct the tree:

```
[Exp.If = if [Exp.Int = INT]]
```

Because the production used to construct the tree appears in the rightmost set of contextual tokens of the contextual symbol Exp$^{\{Exp.If\}}$ in the grammar, the tree is constructed with the sets of leftmost and rightmost tokens containing Exp.If.

As the parser can still perform another reduce action from configuration (II), it does so, reaching configuration (III). Using the tree for the if expression, SGLR creates the following tree, by reducing using an Exp.Add production:

```
[Exp.Add = [Exp.Int = INT] + [Exp.If = if [Exp.Int = INT]]]
```

Note that since the tree for the if expression is used as the rightmost tree when applying the reduce action, and this tree has a non-empty set of rightmost contextual tokens, the set is propagated when creating the tree for the addition.

After shifting the additional symbols from the input, and performing a reduce action that creates the last [Exp.Int = INT] tree, the parser reaches configuration (IV). At this point, there is no other symbol to shift, so only reduce actions are left to be performed. When reducing using an Exp.Add production, there are two possible paths from the state at the top at the stack. The first path, at the top of the graph, creates a tree corresponding to the addition of two integers, which does not contain any deep priority conflict. The second path, at the bottom, contains a conflict since the set of rightmost tokens for the first tree intersects with the rightmost set of contextual tokens for the contextual symbol Exp$^{\{Exp.If\}}$ in the Exp.Add production. Thus, the data-dependent SGLR uses this information to forbid the reduce action on the path at the bottom. By doing that, it produces only a single tree, solving the deep priority conflict.

### 3.3 Performance Optimizations

As shown in the algorithm for the data-dependent SGLR, the operations necessary to perform disambiguation of deep priority conflicts consist of set-algebraic operations such as union (for data propagation) and intersection (for constraint checking). To optimize our solution, we first map all productions that occur in the sets of contextual tokens of contextual symbols from the grammar to a bitset, limiting the amount of data that needs to be propagated. Thus, if a certain production belongs to a set of tokens, its corresponding bit is set to 1, or to 0, otherwise. Using this approach, constraint checking and data propagation can be achieved at a very low cost by performing bitwise operations on these bitsets. With such optimization we were able to achieve near zero-overhead when comparing our data-dependent approach and the original SGLR, for programs that do not contain deep priority conflicts, as we will show in the next section.

## 4 Evaluation

In this section, we evaluate our approach of declarative disambiguation for solving deep priority conflicts, by applying it to a corpus of real programs. We are interested in answering the following research questions:

**RQ1** For files that do not contain deep priority conflicts, how much overhead is introduced by data-dependent contextual grammars?

**RQ2** For files that do contain deep priority conflicts, how do data-dependent contextual grammars perform when solving such conflicts, in comparison to related work?

In order to tackle the aforementioned research questions, it is essential to partition a data set in files that are free of deep priority conflicts, and files that are known to have deep priority conflicts. We re-use a corpus of the top-10 trending OCaml and Java projects on GitHub. The corpus was qualitatively analyzed by Souza Amorim, Steindorfer, and Visser [20], listing the types of priority conflicts each file from the projects contains. For both languages we partitioned the files into two groups according to their analysis results: files are free of deep priority conflicts (and therefore can be parsed by parsers without sophisticated disambiguation mechanisms), and files that contain deep priority conflicts. Table 1 lists the projects contained in the corpus, the total number of source files contained in those project, and the (relative) number of files with deep priority conflicts for OCaml and Java, respectively. Based on the research questions listed before, we can formulate our hypotheses:

**H1** Due to our lightweight data-dependent disambiguation, we expect single-digit percentage overhead when parsing files that do not contain deep priority conflicts.

**H2** Since disambiguation by grammar transformation produces up to three times bigger [20] grammars, we expect our lightweight data-dependent disambiguation to perform significantly better (i.e., higher double-digit percentage improvements).

### 4.1 Experiment Setup

The benchmarks were executed on a computer with 16 GB RAM and an Intel Core i7-6920HQ CPU with a base frequency of 2.9 GHz and a 8 MB Last-Level Cache. The software stack consisted of Apple's macOS operating system version 10.13.1 (17B48) and an Oracle's Java Virtual Machine (version 8u121).

To obtain statistically rigorous performance numbers, we adhere to best practices for (micro-)benchmarking on the Java Virtual Machine (JVM) as, for example, discussed in Georges, Buytaert, and Eeckhout [11] and Kalibera and Jones [14]. We measure the execution time of batch-parsing the corpus of OCaml and Java sources with the Java Microbenchmarking Harness (JMH), which is a framework to overcome the pitfalls of (micro-)benchmarking. Since the batch-parsing execution times are expected to be in terms of minutes—rather than microbenchmarks that execute in milliseconds—we configured JMH to perform 15 *single-shot* measurements: i.e., forking a fresh virtual machine 15 times and measuring the total batch-parsing time including cold startup.

Luis Eduardo de Souza Amorim, Michael J. Steindorfer, and Eelco Visser

■ **Table 1** Deep Priority Conflicts in OCaml and Java Corpus.

| OCaml Project | Affected Files | Java Project | Affected Files |
| --- | --- | --- | --- |
| FStar | 6 / 160 (3.8%) | Matisse | 0 / 41 (0.0%) |
| bincat | 5 / 26 (19.2%) | RxJava | 0 / 1469 (0.0%) |
| bucklescript | 85 / 885 ( 9.6%) | aurora-imui | 0 / 55 (0.0%) |
| coq | 158 / 417 (37.9%) | gitpitch | 0 / 45 (0.0%) |
| flow | 52 / 305 (17.0%) | kotlin | 0 / 3854 (0.0%) |
| infer | 33 / 234 (14.1%) | leetcode | 0 / 94 (0.0%) |
| ocaml | 112 / 909 (12.3%) | litho | 0 / 510 (0.0%) |
| reason | 4 / 36 (11.1%) | lottie-android | 0 / 109 (0.0%) |
| spec | 4 / 40 (10.0%) | spring-boot | 2 / 3444 (0.06%) |
| tezos | 71 / 149 (47.7%) | vlayout | 0 / 46 (0.0%) |
| All | 530 / 3161 (16.8%) | All | 2 / 9667 (0.02%) |

For executing the benchmarks, we disabled CPU frequency scaling, disabled background processes as much as possible, and fixed the virtual machine heap sizes to 10 GB for benchmark execution. The benchmark setup was tested and tuned to yield accurate measurements with relative errors of typically less than 2 % of the execution time. We report the measurement error as Median Absolute Deviation (MAD), which is a robust statistical measure of variability that is resilient to small numbers of outliers.

### 4.2 Experiment Results

The results of our experiment are illustrated in Table 2. We first report the precision of the individual data points. For all the data points, the relative measurement errors are in the range of 1.0 % to 4.1 % with a median error of 1.6 %; the absolute amounts are printed in the table next to the benchmark runtimes (cf. column *Time (seconds)*).

**Cost of Disambiguating Deep Priority Conflicts (Hypothesis H1)** Column *Cost* shows how the parser's performance is affected by supporting the disambiguation of deep priority conflicts. The cost measurements were performed solely for the data sets that are guaranteed to be free of deep priority conflicts, since we use a parser without deep priority conflict disambiguation as a baseline. The results show that the cost of disambiguation with data-dependency is between 1 % (OCaml) and 2 % (Java), supporting Hypothesis H1. Note that the result for OCaml is not statistically significant, i.e., the 1 % difference may as well be in the margin of error. For the Java case, the result is statistically significant, however the error intervals are very close and almost overlap. We conclude that Hypothesis H1 is supported by our experiment: the cost of declarative disambiguation is clearly below 10 %.

**Towards Zero-Overhead Disambiguation of Deep Priority Conflicts**

■ **Table 2** Benchmark Results when parsing the OCaml and Java Corpus.

| Language | Data Set | Disambiguation | Time (seconds) | | | Speedup | Cost |
|---|---|---|---|---|---|---|---|
| Java | with conflicts | data-dependent | 0.18 | ± | 0.00 | 1.29× | — |
| | | rewriting | 0.23 | ± | 0.00 | 1.00× | — |
| Java | without conflicts | data-dependent | 270.64 | ± | 1.28 | 1.73× | 1.02× |
| | | rewriting | 467.20 | ± | 4.03 | 1.00× | 1.77× |
| | | none | 264.20 | ± | 2.36 | — | 1.00× |
| OCaml | with conflicts | data-dependent | 80.60 | ± | 1.48 | 1.54× | — |
| | | rewriting | 123.75 | ± | 1.02 | 1.00× | — |
| OCaml | without conflicts | data-dependent | 89.82 | ± | 0.51 | 1.46× | 1.01× |
| | | rewriting | 130.71 | ± | 0.55 | 1.00× | 1.48× |
| | | none | 88.58 | ± | 0.98 | — | 1.00× |

**Data-Dependent Disambiguation versus Grammar Rewriting (Hypothesis H2)** Column *Speedup* of Table 2 shows the performance improvements of data-dependent disambiguation over disambiguation via grammar rewriting (baseline). In all tested configurations, data-dependent disambiguation speeds-up from 1.29 x to 1.73 x, reducing batch parse times considerably. E.g., parse time for the conflict-free Java corpus reduced from 467.20 s to 270.64 s. We conclude that Hypothesis H2 is supported by our experiment: data-dependent disambiguation outperforms disambiguation via grammar rewriting as discussed in Adams and Might [2] and Souza Amorim, Haudebourg, and Visser [19].

### 4.3 Threats to Validity

To counter internal threats to validity, we properly tested the data-dependent implementation and assured that it produces abstract syntax trees identical to the contextual grammars. For the data sets that are guaranteed to be free of deep priority conflicts, we also assured that the resulting parse trees are identical to the trees from the corresponding non-disambiguating grammar. In all scenarios, we checked that each resulting parse tree is indeed free of ambiguities, cross-validating the findings from the empirical pilot study [20] that accompanies the corpus.

To counter external threats to validity, we carefully designed and implemented our approach to use a lightweight form of data-dependency selectively, solely disambiguating deep priority conflicts. The delta to a baseline SGLR parser without support for disambiguation of deep conflicts is minimal: it requires the addition of a few lines of code, as shown in Listing 4. Therefore, we are confident that the observed cost of 1 % to 2 % for disambiguating deep priority conflicts remains steady, even when using different or larger data sets. We are also confident that the significance of the performance improvement remains clearly observable regardless of the used data sets, because it is commonly known that grammar rewriting blows-up the grammars and the resulting parse tables [20], negatively impacting parsing performance. Nevertheless, the size and choice of our corpus arguably remains an external threat to validity.[5]

## 5 Related Work

In the following section, we highlight previous work on disambiguation of conflicts that arise from the declarative specification of operator precedence and associativity in context-free grammars, grouped by the phase *when* disambiguation happens.

### 5.1 Disambiguation by Grammar Rewriting

Ambiguities that arise from operator precedence and associativity can be avoided by rewriting the grammar to an unambiguous one. In the following, we list declarative disambiguation techniques that try to automatically derive unambiguous grammars.

Aasa [1] proposes a grammar rewriting technique that addresses priority conflicts by generating new non-terminals with explicit precedence levels, forbidding the construction of trees that could cause a conflict. The approach addresses shallow conflicts as well as deep conflicts of type operator-style. Due to a restriction—productions may not have overlapping prefixes or suffixes—it cannot solve dangling-else conflicts.

Thorup [21] presents a grammar transformation algorithm that constructs an unambiguous grammar, given an ambiguous grammar, and a set of counterexamples (i.e., illegal parse trees). The resulting grammar encodes the parse trees bottom-up, while removing grammar symbols that correspond to illegal trees. Thorup's approach specifically supports dangling-else, but does not generalize the construction of counterexamples to capture arbitrary deep priority conflicts.

Conceptually similar to Thorup's idea, Adams and Might [2] propose a grammar rewriting solution, where invalid patterns are expressed using tree automata [9]. Each type of conflict should be expressed as a tree automaton, representing the pattern of the counterexample. Intersecting the counterexample automata with the original context-free grammar yields an unambiguous grammar as result. The authors address all conflicts shown in this paper, with the exception of longest match.[6]

Afroozeh, Brand, Johnstone, Scott, and Vinju [3] describe a *safe* semantics for disambiguation by grammar rewriting that only excludes trees that are part of an ambiguity. While their semantics does cover shallow priority conflicts and deep priority conflicts of type operator-style, it addresses neither dangling-else nor longest match.

---

[5] For the Java data set with conflicts, we would even assume that performance further improves with a larger data set. Unlike the other data sets, the current Java data set with conflicts consists of only two files, because deep conflicts are scarce in Java. The small data set is disadvantaged, because we compare batch-parsing time including cold startup time.

[6] Due to the expressivity of tree automata, we assume that longest match could be supported.

Contextual grammars [19] generalize the approach by Afroozeh, Brand, Johnstone, Scott, and Vinju [3], by supporting a *"safe"* semantics for disambiguation of arbitrary deep priority conflicts. The authors analyze and address the root causes of deep priority conflicts. Their grammar analysis yields as a result, combinations of conflicting productions that may rise to a deep priority conflict. The authors show that deep conflicts can only occur in specific paths in the parse trees. Illegal patterns are conceptually described as deep pattern matches, and implemented by means of recursive grammar rewriting that forbids invalid trees to be constructed. Rewriting is used solely for solving deep priority conflicts; disambiguation of shallow conflicts happens at parse table generation.

All related work mentioned above suffers from the same performance issues: large unambiguous grammars as a result of recursive rewriting, with even larger parse tables that have a low-coverage of parsing states when parsing programs [20]. By contrast, our lightweight data-dependent disambiguation technique avoids grammar transformations and is able to reuse LR parse tables of grammars that do not solve deep priority conflicts, resulting in high parse table coverage and a low overhead of 1 % to 2 % for disambiguation. By reusing the grammar analysis results of contextual grammars, but implementing our mechanism for disambiguation via lightweight data-dependency, we are able to handle arbitrary deep priority conflicts, including deep conflicts caused by indirect recursion.

## 5.2 Disambiguation at Parser Generation

Instead of changing the original grammar, some techniques perform disambiguation of priority conflicts at parser generation time. Even though these solutions do not require changing the productions of the original grammar, they might still be restricted to a certain parser, e.g., by depending on specific characteristics of a parsing technique, or by requiring modifications in the parser generation algorithm.

YACC [13] resolves ambiguities concerning operator precedence and associativity by solving shift/reduce conflicts that occur in LR parse tables. Even though the decision is made dynamically depending on the current lookahead token, the grammar has to specify the default action to take in the conflicting state, given the precedence of the operators involved in the conflict specified in the grammar. This technique has two major drawbacks. First, users have to reason in terms of shift/reduce conflicts, and annotate the grammar if a shift action should be preferred over reduce, or vice versa, to resolve conflicts. Second, YACC's disambiguation does not apply to scannerless parsers, since it requires knowing the lookahead token for decision making. More specifically, YACC's disambiguation does not apply to any parser that relies on character-level grammars [17, 18, 25], which are useful to avoid issues when composing grammars of different languages [6, 7].

Klint and Visser [15] propose a semantics for declarative disambiguation based on disambiguation filters. This semantics has been implemented by SDF2, so invalid tree patterns can be constructed from SDF2 priority declarations [23], and used in a disambiguation filter. The implementation relies on a custom LR parse table generator, as SDF2 parse tables encode goto transitions between states using productions, forbidding transitions that could construct an invalid tree according to the tree patterns.

Because this semantics only targets conflicts by checking a parent-child relation in a tree, this solution is not able to solve deep priority conflicts.

### 5.3 Disambiguation while Parsing

Ambiguities from operator precedence may also be addressed at parse time. Such approaches have the advantage of using the original (or a slightly modified) grammar as input, but require adaptations of the parsing algorithm that may cause overhead when parsing programs that are free of priority conflicts.

Afroozeh and Izmaylova [5] introduce a solution for disambiguating priority conflicts on the basis of a full-fledged data-dependent grammar formalism. The authors implemented their approach in a generalized LL parser named Iguana [4]. Iguana addresses shallow priority conflicts and operator-style deep conflicts using data-dependent grammars, but the approach does not extend to dangling-else nor longest match. Given the experimental setup described in their paper [5], it is not possible to assess the overhead of solving deep priority conflicts, because no analysis has been performed on programs free of deep priority conflicts. When solving the shallow conflicts present in the Java 7 grammar, Afroozeh and Izmaylova's approach causes, on average, 5 % overhead compared to the unambiguous Java grammar that directly encodes precedence and associativity. In contrast, in this paper we measure the cost of disambiguating deep priority conflicts by parsing programs that are known to be free of deep priority conflicts. Our lightweight data-dependent solution has negligible overhead to solve deep priority conflicts, furthermore, it is able to address more types of deep priority conflicts than Iguana.

The ALL(*) parsing algorithm of ANTLR [16] also handles operator precedence dynamically by means of semantic predicates. Because top-down parsers cannot handle left-recursive rules, the grammar is first rewritten to eliminate direct recursion using a technique known as *precedence climbing* [8]. Next, semantic predicates that are evaluated at parse time may filter invalid trees according to the order in which productions are defined in the grammar. The predicates are interwoven in the grammar productions representing the constraints to avoid producing invalid trees. In our case, the constraints are encoded in contextual non-terminals, as they indicate the trees that a non-terminal should not produce as its leftmost or rightmost child. Furthermore, we assume that the ANTLR solution to disambiguate deep conflicts have a bigger impact on performance than our lightweight data-dependency, more specifically when parsing programs without conflicts, as it uses similar techniques to data-dependent grammars.

Finally, Erdweg, Rendel, Kästner, and Ostermann [10] implemented a disambiguation strategy at parse time for SGLR, to support layout-sensitive languages. The disambiguation mechanism consists of propagating information about layout when constructing the parse trees, and enforcing constraints that are defined as attributes of productions in SDF grammars. While we only propagate information about leftmost and rightmost subtrees, their approach needs to propagate line and column positions of all terminal symbols that were used to construct a tree. In our case, we express constraints using sets of contextual symbols, checking them using set-algebraic operations. By using an optimized bitset representation, our approach achieves near zero-overhead when disambiguating deep priority conflicts. In contrast, Erdweg, Rendel, Kästner, and Ostermann state that their layout-sensitive parsing approach is practicable with an average slowdown of 1.8 x compared to a layout-insensitive solution. The authors mix enforcing constraints at parse-time and post-parse, compared to our solution that solely disambiguates deep priority conflicts at parse time.

## 6   Conclusions

In this paper, we presented a novel low-overhead implementation technique for disambiguating deep priority conflicts with data-dependency. The approach was implemented in a scannerless generalized LR parser, and evaluated by benchmarking parsing performance of a corpus of popular Java and OCaml projects on Github. Results show that our data-dependent technique cuts down the cost of disambiguating deep priority conflicts to 1 % to 2 %, improving significantly over contextual grammar rewriting strategies that have an overhead of 48 % to 77 %, as shown in Section 4. By using data-dependency selectively for just solving deep priority conflicts, we were able to reuse the (compact) LR parse tables of grammars that do not disambiguate deep conflicts, avoiding the typical problems of parse-table blowup of grammar rewriting strategies. Overall, we showed that declarative disambiguation can indeed be solved with almost no cost.

**Acknowledgements**   The work presented in this paper was partially funded by CAPES (Coordenação de Aperfeiçoamento de Pessoal de Nível Superior - Brazil) and by the NWO VICI *Language Designer's Workbench* project (639.023.206). We would also like to thank Jasper Denkers for his work on JSGLR2, and Peter Mosses and the anonymous reviewers for their detailed feedback.

**About the authors**

**Luis Eduardo de Souza Amorim** is a PhD student working at the Programming Languages/Software Technologies group at Delft University of Technology. His work focuses on syntax definition formalisms and parsing. Eduardo received degrees in computer science from Universidade Federal de Viçosa, in Brazil (BSc 2011, MSc 2013). Contact him at l.e.desouzaamorim-1@tudelft.nl.

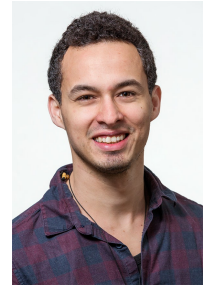

**Michael J. Steindorfer** is a senior software engineer working in industry and a guest researcher at the Delft University of Technology. His research and engineering efforts focus on optimizing functional data structures, the design and implementation of programming languages, and improving big data processing runtimes for cloud infrastructures. You can contact him at michael@steindorfer.name.

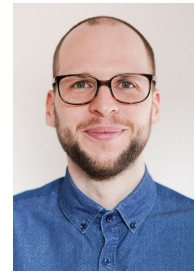

**Eelco Visser** is Antoni van Leeuwenhoek Professor of Computer Science and chair of the Programming Languages Group at Delft University of Technology. His current research is on the foundation and implementation of declarative specification of programming languages. Contact him at e.visser@tudelft.nl, and find further information at https://eelcovisser.org.

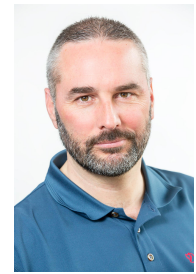